\documentclass[prl,twocolumn,showpacs,amsmath,amssymb]
{revtex4}

\usepackage[dvips]{color}
\usepackage{graphicx}

\begin{document}

\title{Collective Oscillations of Strongly Correlated 
One-Dimensional Bosons on a Lattice}

\author{M. Rigol}
\affiliation{Physics Department, University of California, Davis,
CA 95616, USA}
\author{V. Rousseau}
\affiliation{Physics Department, University of California, Davis,
CA 95616, USA}
\author{R. T. Scalettar}
\affiliation{Physics Department, University of California, Davis,
CA 95616, USA}
\author{R. R. P. Singh}
\affiliation{Physics Department, University of California, Davis,
CA 95616, USA}

\begin{abstract}
We study the dipole oscillations of strongly correlated 1D bosons, 
in the hard-core limit, on a lattice, by an exact numerical approach. 
We show that far from the regime where a Mott insulator 
appears in the system, damping is always present and increases
for larger initial displacements of the trap, 
causing dramatic changes in the momentum distribution, $n_k$. 
When a Mott insulator sets in the middle of the trap, the 
center of mass barely moves after an initial displacement, 
and $n_k$ remains very similar to the one in the ground 
state. We also study changes introduced by the damping in the 
natural orbital occupations, and the revival of the 
center of mass oscillations after long times.
\end{abstract}

\maketitle

Experimental realization of one-dimensional (1D) 
atomic gases \cite{moritz03,tolra04} has opened the possibility 
of studying equilibrium and non-equilibrium properties
of systems in which quantum fluctuations play a fundamental 
role. For example, the 1D superfluid--Mott-insulator 
transition \cite{batrouni02} was observed experimentally by 
St\"oferle {\it et~al.} \cite{stoferle03}. In addition, 
1D geometries have allowed the observation of the strongly 
correlated hard-core boson limit with \cite{paredes04} 
and without \cite{kinoshita04} a lattice along the 1D tubes.

Recent experiments studying transport properties of 
trapped 1D Bose gases \cite{stoferle03,fertig04} have 
reported a large damping of the center-of-mass (c.m.) 
oscillation when an axial lattice is present in the 
system. Additionally, an overdamping was observed in which 
the c.m. remains displaced from the middle of the trap
\cite{stoferle03,fertig04}. In Ref.\ \cite{stoferle03} 
this effect was related to the presence of an incompressible 
Mott insulating phase, while Fertig {\it et~al.} \cite{fertig04} 
reported to be far from this regime. For weak interactions, 
and small trap displacements, the damping has been argued
(within mean field approximations) to relate to the existence 
of non-condensate fractions, which cause a ``dissipative'' 
behavior of the c.m. oscillations \cite{polkovnikov04,banacloche04}. 
For large trap displacements, dynamical instabilities 
\cite{wu01,smerzi02} can also produce damping.

For strongly correlated 1D bosons on a lattice, the 
mean-field description assuming Bose-Einstein Condensation 
(BEC) breaks down. For hard-core bosons (HCB), the largest 
eigenvalues of the one-particle density matrix (OPDM) 
[also called natural orbitals (NO)] scale proportional 
to $\sqrt{N_b}$ [$N_b$ is the number of bosons]. 
There is no BEC. However, there is a finite superfluid fraction 
$\rho_s$, which for the periodic case (no trap) can be easily 
calculated $\rho^{1D}_s=\dfrac{\sin(\pi N_b/N)}{\pi N_b/N}$ , 
where $N$ is the total number of lattice sites
(see {\it e.g.}\ Ref.\ \cite{laflorencie01}). In two dimensions 
(2D) HCB exhibit BEC at zero temperature. However, the 2D superfluid
fraction is smaller, at any density, than in 1D, though larger 
than the 2D condensate fraction \cite{bernardet02}. 
These results exemplify the non-trivial differences between BEC 
and superfluidity. It is unclear how much the lack of a true BEC 
is central to understanding transport properties in 1D.

The presence of the trap in the experiments with ultracold 
quantum gases generates an inhomogeneous density profile \cite{batrouni02,rigol04_1}, and modifies the single particle 
spectrum of the periodic lattice \cite{rigol03_3}. This could,
in general, change the behavior of the superfluid and condensate fractions.
However, for the 1D HCB, the power-law decay of the OPDM 
($\rho_x\sim 1/\sqrt{x}$), and the occupation of the lowest 
NO ($\lambda\sim \sqrt{N_b}$), have been shown to be the same as
in the periodic system \cite{rigol04_1}. 

In this work we study the dynamics of the c.m. oscillations of 
hard-core bosons (HCB) confined in 1D harmonic traps with an 
underlying lattice, by an exact numerical method. 
Our purpose is to examine the damping of 
such oscillations, and its consequences in the momentum distribution
function ($n_k$) and natural orbital (NO) occupation.
We show that far from the 
Mott-insulating (MI) regime in the middle of the trap, the damping 
of the c.m. oscillations grows with initial displacement and in the large
damping regime, causes dramatic changes in $n_k$. 
On the contrary, when the MI appears in the middle of the trap,
even for small initial displacements the c.m. barely moves from its 
initial position, and $n_k$ remains very similar to the one in the 
ground state.

The HCB Hamiltonian can be written as 
\begin{equation}
\label{HamHCB} H = -t \sum_{i} \left( b^\dagger_{i} b^{}_{i+1}
+ h.c. \right) + V_2 \sum_{i} x_i^2 \ n_{i },
\end{equation}
with the addition of the on-site constraints
$b^{\dagger 2}_{i}= b^2_{i}=0$, 
$\left\lbrace  b_{i},b^{\dagger}_{i}\right\rbrace =1$.
These constraints on the creation ($b^{\dagger}_{i}$) and 
annihilation ($b_{i}$) operators avoid double or higher occupancy.  
In Eq.\ (\ref{HamHCB}), the hopping parameter is denoted by $t$, 
and the last term describes a harmonic confining potential
with curvature $V_{2}$. $n_{i}= b^{\dagger}_{i}b_{i}$ is the 
particle number operator. The non-equilibrium dynamics 
of the system, when at time $\tau=0$ the trap is displaced a 
distance $x_0$, is obtained by means of the exact approach 
introduced in Ref.\ \cite{rigol04_2}. It is based on the 
Jordan-Wigner transformation, and allows obtaining the 
HCB one-particle Green's function from which all the
quantities analyzed in this work are calculated. In addition
to being exact, this method allows studying relatively large 
system sizes.

Previous studies have shown that the key parameter that controls
the thermodynamic behavior of this model is the characteristic density
$\tilde{\rho}=N_b \sqrt{V_2/t}$ \cite{rigol04_1}. 
As $\tilde{\rho}\to 0$, one recovers the continuum limit. 
On the other hand, as $\tilde{\rho}$ increases
beyond a critical value ($\tilde{\rho}_c\sim 2.6-2.7$) a MI 
region builds up in the middle of the trap, 
where the density is pinned to unity with no quantum fluctuations.

In Fig.\ \ref{CMvsDisp}(a)-(c) we show the c.m. oscillations 
of a trapped HCB system for $\tilde{\rho}=1<\tilde{\rho}_c$, 
far from the regime with a MI in the middle of the 
trap. At time $\tau=0$ the maximum density in the trap is $n=0.48$, 
and the cloud 
radius is $R_0\sim 145 a$, with $a$ the lattice constant. 
Fig.\ \ref{CMvsDisp}(a) shows that even for a very small 
displacement, $7\%$ of the cloud radius, a damping
of the oscillations occurs. For this displacement the 
maximum momentum of the c.m. of the system is $\sim \pi/30a$,
very far from the momentum at which the classical 
modulation instability occurs, which for HCB is $\pi/2a$
(see {\it e.g.}\ Ref.\ \cite{polkovnikov04a}). Since HCB can 
be mapped into noninteracting fermions, one can immediately 
see that damping occurs for any initial density 
and trap displacement. In the fermion language, the damping is caused
by a dephasing of the particles, which arises due to their dispersion
in the lattice-harmonic trap system \cite{rigol03_3,pezze04}. As the initial 
displacement of the trap is increased [Fig.\ \ref{CMvsDisp}(b)], 
the damping also increases. For large displacements, Bragg scattering 
starts to occur producing a shift of the c.m. oscillations 
[Fig.\ \ref{CMvsDisp}(c)]. A large damping of the c.m. oscillations can 
also be observed. In Fig.\ \ref{CMvsDisp}(c) the maximum momentum of 
the c.m. is $\sim \pi/4a$, still smaller than $\pi/2a$.
\begin{widetext}
\begin{center}
\begin{figure}[h]
\begin{center}
\includegraphics[width=0.9\textwidth,height=0.42\textwidth]
{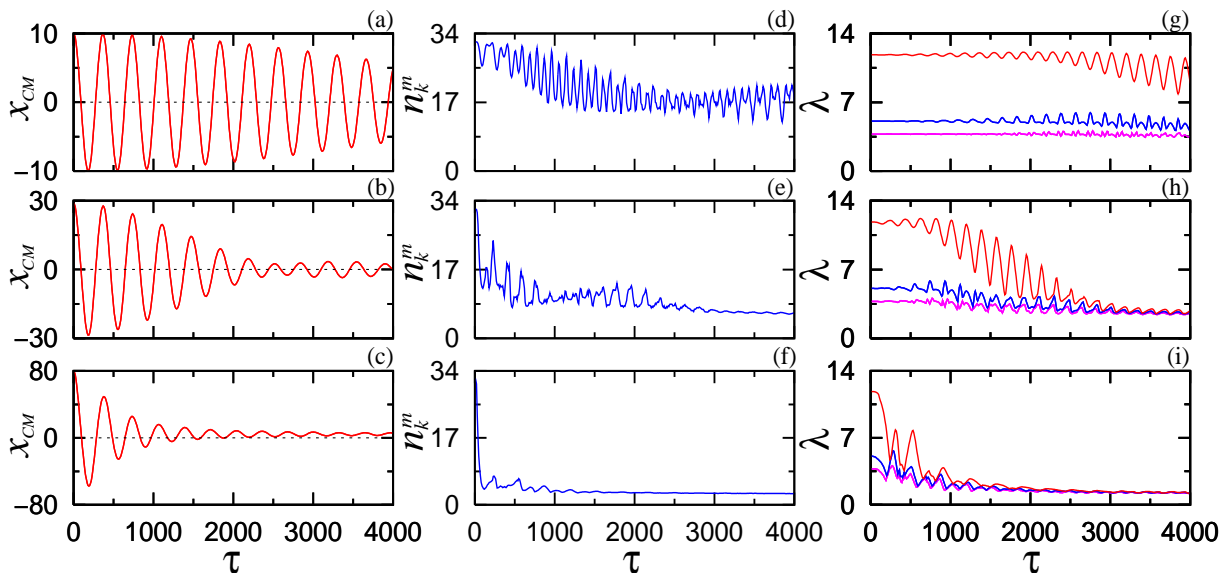}
\end{center}\vspace{-0.3cm}
\caption{(color online). Evolution of the c.m. position (a)-(c), 
maximum value of $n_k$ (d)-(f), 
and the occupations of the three lowest NO (g)-(i), 
vs time ($\tau$, in units of $\hbar/t$). 
The initial trap displacements are 
$x_0=10a$ [$x_0/R_0\sim0.07$] (a),(d),(g), 
$x_0=29a$ [$x_0/R_0\sim0.2$] (b),(e),(h), and 
$x_0=80a$ [$x_0/R_0\sim0.55$] (c),(f),(i).
These systems have $N_b=101$, and $V_2 a^2=10^{-4} t$.}
\label{CMvsDisp}
\end{figure}
\end{center}
\end{widetext}

\begin{figure}[h]
\begin{center}
\includegraphics[width=0.43\textwidth,height=0.27\textwidth]
{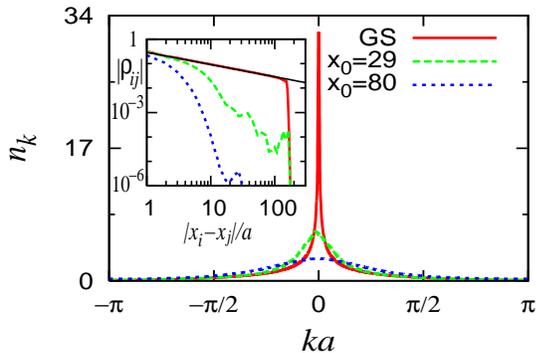}
\end{center} \vspace{-0.6cm}
\caption{(color online). $n_k$ of the ground state (GS) compared 
with the ones at $\tau=4000\hbar/t$ for two different initial displacements
of the trap ($N_b=101$, and $V_2 a^2=10^{-4} t$). 
As shown in Figs.\ \ref{CMvsDisp}(b),(c), at $\tau=4000\hbar/t$ 
the motion of the CM is totally damped. The inset shows the OPDM 
for the same systems, the straight line is 
$\rho_{ij}\sim 1/\sqrt{|x_i-x_j|/a}$.}
\label{perfilK}
\end{figure}

We next analyze the consequences of the damping of the c.m. 
motion in another physical observable, the momentum distribution 
function $n_k=(a/\zeta)\sum_{jl} e^{-ik(j-l)}\rho_{jl}$, with 
$\zeta=\left( V_2/t\right)^{-1/2}$  \cite{rigol04_1}, and
$\rho_{jl}=\langle b_j^\dagger b_l\rangle$ the OPDM. 
In Fig.\ \ref{CMvsDisp}(d)-(f) we show how the maximum
value of $n_k$ ($n_k^m$) evolves as a function of time. 
In Fig.\ \ref{CMvsDisp}(d) one can 
see that even for small damping, the oscillations of $n_k^m$ are 
accompanied by an overall decrease of its mean value, which reaches 
almost half of the original height for $\tau=4000 \hbar/t$. The changes in
$n_k^m$ become more dramatic with increasing $x_0$. In the 
large damping regime [Fig.\ \ref{CMvsDisp}(f)], one can see that 
$n_k^m$ reduces almost to its minimum value in the first oscillation
of the c.m. The reduction of $n_k^m$ is accompanied by a large increment
in the full width at half maximum $w$, as shown in Fig.\ \ref{perfilK}.
These results are in contrast with the ones for the equivalent 
noninteracting fermions \cite{rigol03_3}, where after the damping 
of the c.m. oscillations, $n_k^m$ and $w$ are almost the same as 
the ones at $\tau=0$. (Contrary to the density, the evolution of $n_k$ for 
fermions and bosons is very different.)

At this point it is relevant to know the behavior 
of the ``condensate'' occupation during the oscillations of the c.m. 
For these systems it is important not to confuse $n_k^m$ with the 
``condensate'' occupation. The condensate can be defined as the largest 
eigenvalue of the OPDM (the lowest NO) \cite{penrose56,leggett01}, 
and since for HCB its occupation $\lambda_0$ scales proportionally 
to $\sqrt{N_b}$ \cite{rigol04_1,forrester03}, we prefer to call 
it a quasicondensate ($\lambda_0\rightarrow\infty$ for 
$N_b\rightarrow\infty$, but $\lambda_0/N_b\rightarrow 0$).
One particular example in which the differences between $n_k^m$
and $\lambda_0$ are extreme was reported by one of us in 
Ref.\ \cite{rigol04_3}. During the expansion of the HCB gas,
after turning the harmonic trap off, $n_k^m$ decreases
while $\lambda_0$ increases.

In Fig.\ \ref{CMvsDisp}(g)-(i) we show the time evolution of the 
three lowest NO occupations (the highest occupied ones). It 
can be seen that indeed the behavior of $\lambda_0$ is different 
from the one of $n_k^m$. At short times, and for low damping rates 
[Fig.\ \ref{CMvsDisp}(g)], the lowest NO {\em occupation} does not 
change during the oscillations of the system, even though the 
lowest NO itself does oscillate in a coherent manner, as would 
be the case for an ideal BEC when there is no lattice. Only when 
the c.m. energy is transfered into excitation energy,
the lowest NO occupations (and the shape of their wave-functions) 
start to change. This reflects the onset of a cut-off to the 
off-diagonal quasi-long range order initially present in the
OPDM, and it is similar to the appearance of a finite temperature
in the system. Such effect can be seen in the inset in 
Fig.\ \ref{perfilK} where we show the OPDM of the ground state
when the c.m. oscillations are damped. 
In the latter cases, the fast decays of the OPDM reflected 
in the plots are very close to exponentials. As expected,
the largest damping produces the fastest decay.
The consequent reduction of the lowest NO occupations 
can be seen in Fig.\ \ref{CMvsDisp}(h),(i).

\begin{figure}[h]
\begin{center}
\includegraphics[width=0.40\textwidth,height=0.428\textwidth]
{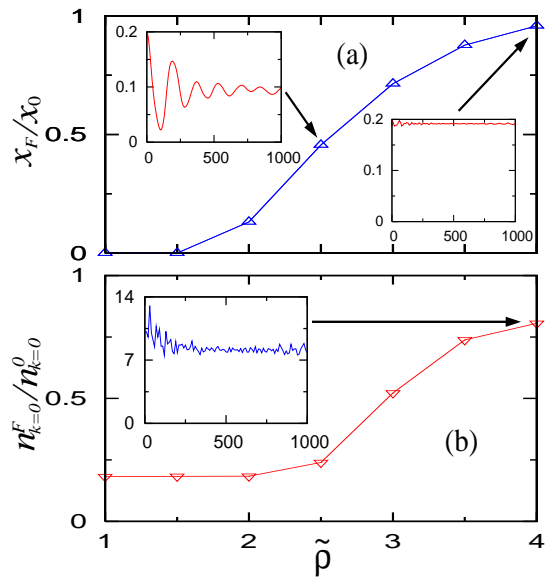}
\end{center} \vspace{-0.6cm}
\caption{(color online). (a) Ratio between the c.m. position after 
damping ($x_F$) and the initial trap displacement ($x_0$) vs $\tilde{\rho}$.
The insets show $x_{CM}/R_0$ vs $\tau$ for the signaled $\tilde{\rho}$. 
(b) Ratio between $n_{k=0}$ after damping ($n^F_{k=0}$) and $n_{k=0}$ 
at $\tau=0$ ($n^0_{k=0}$) vs $\tilde{\rho}$. The inset shows 
$n_k^m$ vs $\tau=0$ for the signaled $\tilde{\rho}$.}
\label{CMvsRho}
\end{figure}

\begin{figure}[h]
\begin{center}
\includegraphics[width=0.47\textwidth,height=0.20\textwidth]
{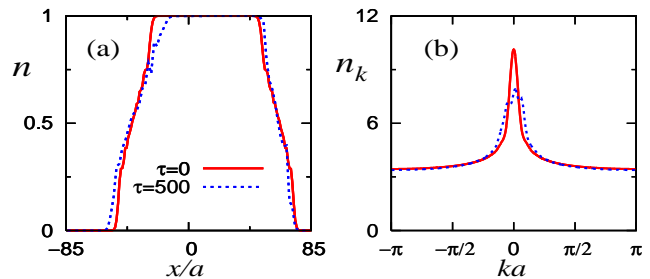}
\end{center} \vspace{-0.6cm}
\caption{(color online). Density and momentum profiles for a 
system with $N_b=101$, and $\tilde{\rho}= 4$. See also 
results for $\tilde{\rho}= 4$ in the insets of Fig.\ \ref{CMvsRho}.}
\label{Perfil}
\end{figure}

So far we have analyzed trapped systems with no MI phase. 
In what follows we study the consequences of approaching the 
MI regime on the damping of the c.m. oscillations. 
Fig.\ \ref{CMvsRho}(a) shows that after damping, when the 
characteristic density of the system is increased, 
the final c.m. position is displaced from the center of the trap. 
We kept the ratio between the initial trap displacement $x_0$ 
and the cloud radius $R_0$ constant, and equal to $0.2$ as in 
Figs.\ \ref{CMvsDisp}(b),(e),(h). The insets show the results 
for $x_{CM}/x_0$ vs $\tau$ in traps with $\tilde{\rho}= 2.5$, 
just before the MI phase appears, and for $\tilde{\rho}=4$, 
when a MI domain is present in the center of the trap 
[Fig.\ \ref{Perfil}(a)]. These results show that 
(i) the damping rate increases with increasing $\tilde{\rho}$, 
and Bragg scattering keeps the c.m. displaced from the middle 
of the trap, (ii) when there is a MI domain ($\tilde{\rho}=4$), 
the c.m. barely displaces from its original position. 
These results are in agreement with the experiments reported 
by Stoferle {\it et.~al} \cite{stoferle03}, and differ from the 
ones in the weakly interacting regime where for small initial 
trap displacements no shift of the c.m. position is observed 
after damping \cite{polkovnikov04,banacloche04}.

One interesting feature that appears with the formation of the MI 
can be seen in the inset in Fig.\ \ref{CMvsRho}(b). 
Although $n_k^m$ at $\tau=0$ is small compared with $n_k^m$ 
for smaller $\tilde{\rho}$, its value remains almost the same 
during the evolution of the system. The same occurs with $w$,
the width of $n_k$, as seen in Fig.\ \ref{Perfil}(b). The above result 
can be intuitively understood by the fact that almost no c.m. energy 
has been converted into excitation energy, which keeps the system in 
a state with $n_k$, and NO occupations, similar to the ones 
in the ground state. Something similar may occur in experiments 
in the overdamped regime, when the c.m. almost does not change 
from its initial position \cite{fertig04}. However, we should remark 
that we do obtain big changes of $n_k^m$, and $w$, [Figs.\ \ref{perfilK} 
and \ref{CMvsRho}(b)] as a consequence of the damping in systems 
with no MI. This is in contrast with the behavior of $w$ one  
infers from the constant cloud widths after time-of-flight (TOF) 
reported experimentally in Ref.\ \cite{fertig04}.
The reason for this difference may be that TOF measurements, 
due to interparticle interactions during the expansion, 
may not give the original $n_k$ in the trap. 
(An explicit example was presented in Ref.\ \cite{rigol04_3}.)
If this is the case, other experimental techniques will be 
needed to determine the effects of the damping on $n_k$.

\begin{figure}[h]
\begin{center}
\includegraphics[width=0.4\textwidth,height=0.385\textwidth]
{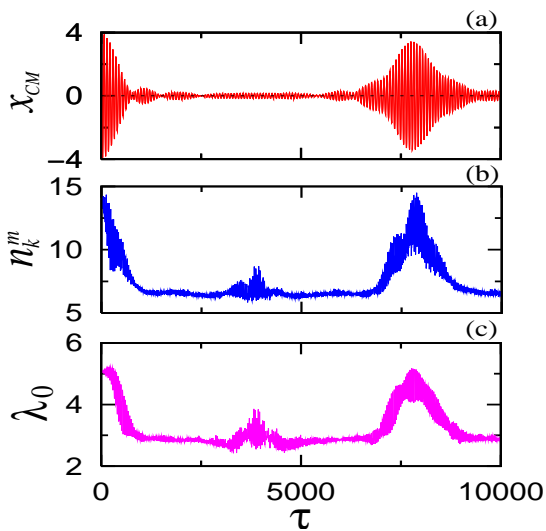}
\end{center}\vspace{-0.6cm}
\caption{(color online). Long time evolution of the c.m. 
position (a), maximum value of $n_k$ (b), 
and the occupation of the lowest NO (c), 
vs time ($\tau$, in units of $\hbar/t$). 
The initial trap displacement is 
$x_0=4a$ [$x_0/R_0\sim 0.1$], and the system has 
$N_b=20$, and $\tilde{\rho}=1.0$.}
\label{CM_Recur}
\end{figure}
One final remark is in order concerning the damping discussed here.
Since the experimental systems with ultracold quantum gases 
are almost closed systems, damping has a different meaning 
when compared with systems in contact with a reservoir, 
where the energy is truly dissipated. In the systems we 
have analyzed, the energy is {\it conserved}. In Fig.\ \ref{CM_Recur} 
we show a long time evolution of the c.m. oscillations, $n_k^m$, 
and the lowest NO occupation, after an initial displacement of the 
center of the trap. As seen in these figures, a revival of 
the c.m. oscillations occurs, with an increase of $n_k^m$ and 
$\lambda_0$ to values similar to the ones at $\tau=0$. This 
suggests that ultracold gases experiments could be used to 
study quantum Poincare recurrences \cite{bocchieri57}. 
The relation between long time dynamics in integrable systems,
such as HCB analyzed here, and nonintegrable (finite on-site 
repulsion) models, is an important open question (of broad 
relevance to condensed matter physics) that needs 
further analysis.

In summary, we have presented a detailed, and exact, study of the 
damping of the c.m. oscillation in a system of strongly correlated 
bosons. We have shown that far from the Mott-insulating regime 
large damping occurs only for large initial displacements of the trap, 
and it reduces dramatically the maximum value of $n_k$ and increases 
its full width. The damping also destroys the quasi-long range 
order of the OPDM, reducing the occupation of the lowest NO, 
analogous to an increase in the temperature of the system. 
When the Mott insulator is present in the trap, 
overdamping occurs even for small initial displacements. 
In this case the c.m. remains very close to its initial position, 
and no big changes occur in $n_k$. Finally, we have shown that since 
the system is closed, after some time a revival of the c.m. oscillations 
occurs, and $n^m_k$ and $\lambda_0$ also regain their initial values. 

{\it Note added.} Complementary to this work, a related study has 
been recently presented in Ref.\ \cite{rey05}.

We thank G.G. Batrouni, T. Bryant, V. Dunjko, A. Muramatsu, 
M. Olshanii, and D. Thouless for insightful discussions. 
This work was supported by NSF-DMR-0312261 and NSF-DMR-0240918.


\begin{thebibliography}{99}

\vspace{-0.3cm}

\bibitem{moritz03} H. Moritz {\it et~al.}, 
Phys. Rev. Lett. {\bf 91}, 250402 (2003).
 
\bibitem{tolra04} B. L. Tolra {\it et~al.}, 
Phys. Rev. Lett. {\bf 92}, 190401 (2004).

\bibitem{batrouni02} G.~G. Batrouni {\it et~al.}, 
Phys. Rev. Lett. {\bf 89}, 117203 (2002).

\bibitem{stoferle03} T. St\"oferle {\it et~al.}, 
Phys. Rev. Lett. {\bf 92},  130403  (2004).

\bibitem{paredes04} B. Paredes {\it et~al.}, 
Nature (London) {\bf 429}, 277 (2004).

\bibitem{kinoshita04} T. Kinoshita, T. Wenger, and D. S. Weiss, 
Science {\bf 305}, 1125 (2004).

\bibitem{fertig04} C. D. Fertig {\it et~al.}, 
Phys. Rev. Lett. {\bf 94}, 120403 (2005).

\bibitem{polkovnikov04} A. Polkovnikov and D.-W. Wang, 
Phys. Rev. Lett. {\bf 93}, 070401 (2004).

\bibitem{banacloche04} J. Gea-Banacloche {\it et~al.},
cond-mat/0410677.

\bibitem{wu01} B. Wu and Q. Niu, 
Phys. Rev. A {\bf 64}, 061603(R) (2001).

\bibitem{smerzi02} A. Smerzi {\it et~al.}, 
Phys. Rev. Lett. {\bf 89}, 170402 (2002).

\bibitem{laflorencie01} N. Laflorencie, S. Capponi, and 
E. S. S\o{}rensen, Eur. Phys. J B {\bf 24}, 77 (2001).

\bibitem{bernardet02} K. Bernardet {\it et~al.}, 
Phys. Rev. B {\bf 65}, 104519 (2002).

\bibitem{rigol04_1} M. Rigol and A. Muramatsu, 
Phys. Rev. A {\bf 70}, 031603(R) (2004); 
{\bf 72}, 013604 (2005).

\bibitem{rigol03_3} M. Rigol and A. Muramatsu, 
Phys. Rev. A {\bf 70}, 043627 (2004).

\bibitem{rigol04_2} M. Rigol and A. Muramatsu, 
Phys. Rev. Lett. {\bf 93}, 230404 (2004).

\bibitem{polkovnikov04a} A. Polkovnikov {\it et~al.},
Phys. Rev. A {\bf 71}, 063613 (2005).

\bibitem{pezze04} L. Pezz\`e {\it et~al.}, 
Phys. Rev. Lett. {\bf 93}, 120401 (2004).

\bibitem{penrose56} O. Penrose and L. Onsager, 
Phys. Rev. {\bf 104}, 576 (1956).

\bibitem{leggett01}
A. J. Leggett, Rev. Mod. Phys. {\bf 73},  307  (2001).

\bibitem{forrester03} P. J. Forrester {\it et~al.}, 
Phys. Rev. A {\bf 67}, 043607 (2003).

\bibitem{rigol04_3} M. Rigol and A. Muramatsu, 
Phys. Rev. Lett. {\bf 94}, 240403 (2005).

\bibitem{bocchieri57} P. Bocchieri and A. Loinger, 
Phys. Rev. {\bf 107}, 337 (1957); L. S. Schulman,
Phys. Rev. A {\bf 18}, 2379 (1978).

\bibitem{rey05} A. M. Rey {\it et~al.},
cond-mat/0503477.

\end{thebibliography}
\end{document}